
\input phyzzx
\nonstopmode
\sequentialequations
\twelvepoint
\nopubblock
\tolerance=5000
\overfullrule=0pt

\REF\turok{N. Turok, {\it Phys. Rev. Lett.} {\bf 63},
2625 (1989).}

\REF\review{N. Turok, Princeton Report No. PUPT-90-1230 (1990).}

\REF\micro{N. Turok and D. Spergel, {\it Phys. Rev. Lett.} {\bf 64},
2736 (1990).}

\REF\kibble{T. Kibble, {\it J. Phys.} {\bf A9}, 1387 (1976).}

\REF\gs{S. Giddings and A. Strominger, {\it Nucl. Phys.} {\bf B307},
854 (1988).}

\REF\coleman{S. Coleman, {\it Nucl. Phys.} {\bf B310}, 643 (1988).}

\REF\gilbert{G. Gilbert, {\it Nucl. Phys.} {\bf B328}, 159 (1989). }

\REF\nohair{See, for example, B. Carter in {\it General
Relativity.  An Einstein Centenary Survey,} edited by
S.~W.~Hawking and W.~Israel (Cambridge University Press,
Cambridge, 1979).}

\REF\hawking{S. W. Hawking, {\it Commun. Math. Phys.} {\bf 43},
199 (1975).}

\REF\spergel{D. Spergel, private communication.}

\REF\kw{L. Krauss and F. Wilczek, {\it Phys. Rev. Lett.}
{\bf 62}, 1221 (1989).}

\REF\amr{For a review see: M. Alford and J. March-Russell,
{\it Int. J. Mod. Phys.} {\bf B5}, 2641 (1991).}

\REF\non{M. Alford, S. Coleman, and J. March-Russell, {\it Nucl.
Phys.} {\bf B351}, 735 (1991); M. Alford, J. March-Russell, and
F. Wilczek, {\it Nucl. Phys.} {\bf B337}, 695 (1990); J.
Preskill, L.~Krauss, {\it Nucl. Phys.} {\bf B341}, 50 (1990).}

\REF\string{J. H. Schwarz, Caltech Report No. CALT-68/1688
(1990).}

\REF\dst{R. Davis, D. Spergel, and N. Turok, work in progress.
We thank R. Davis and D. Spergel for informing
us of this work.}

\def\fun#1#2{\lower3.6pt\vbox{\baselineskip0pt\lineskip.9pt
  \ialign{$\mathsurround=0pt#1\hfil##\hfil$\crcr#2\crcr\sim\crcr}}}
\def\lap{\mathrel{\mathpalette\fun <}}
\def\gap{\mathrel{\mathpalette\fun >}}
\def\order{{\cal O}}

\def\ts{t_{\rm struc}}
\def\Mp{M_{\rm Pl}}

\let\De=\Delta

\let\la=\lambda

\let\del=\nabla

\let\th=\theta

\let\p=\partial
\let\<=\langle
\let\>=\rangle

\def\comment#1{ \hbox{Comment suppressed here.} }

\line{\hfill IASSNS-HEP-92/6}
\line{\hfill PUPT-92-1304}
\line{\hfill January 1992}
\titlepage
\title{Are Textures Natural?}
\medskip
\author{Marc Kamionkowski\foot{Research supported
by an SSC Fellowship from the Texas National Research
Laboratory Commission. e-mail: kamion@guinness.ias.edu}}
\smallskip
\centerline{{\it School of Natural Sciences}}
\centerline{{\it Institute for Advanced Study}}
\centerline{{\it Olden Lane}}
\centerline{{\it Princeton, N.J. 08540}}
\medskip
\author{John March-Russell\foot{Research supported by NSF
grant
NSF-PHY-90-21984. e-mail: jmr@puhep1.princeton.edu,
jmr@iassns.bitnet}}
\smallskip
\centerline{{\it Joseph Henry Laboratories}}
\centerline{{\it Princeton University}}
\centerline{{\it Princeton, N.J. 08544}}
\medskip

\abstract{We make the simple observation that,
because of global symmetry violating higher-dimension
operators expected to be induced by Planck-scale physics,
textures are generically much too short-lived to be of
use for large-scale structure formation.}

\endpage

The texture scenario for large-scale structure
formation is an attractive alternative to inflation or other,
defect mediated, proposals based on new physics at a high-energy
scale [\turok].  The model assumes the existence of a
nonabelian global continuous symmetry which, when broken at the GUT scale,
leads to a topological defect known as a texture.  With one free
parameter, the symmetry-breaking scale $v$, the model reproduces
galaxy-galaxy correlation functions, finds significant galaxy
clustering on scales of 20-50 $h^{-1}$ Mpc, structure on
larger scales, and coherent velocity fields on scales similar to
those observed in large-scale surveys.  The optimal value of $v$
turns out to be $10^{16}$ GeV, the GUT scale [\review].
Another advantage is that the theory also predicts
a distinctive signature in the cosmic microwave background and
will therefore be testable in the near future [\micro].  We feel
that given the recent attention devoted to, and the astrophysical
promise of the model that an investigation of the
particle-physics behind the model is in order.

In this Letter we point out that the existence of
higher-dimension symmetry-violating operators invalidates the
texture scenario.  We find that the couplings of such operators,
which are expected to be induced at the Planck scale by
quantum-gravity effects, would have to be exponentially small in order
for textures to be effective in generating large-scale
structure.  In other words, in order for textures to work, the
assumed global symmetry must be essentially {\it exact}.  Our argument is
simply that explicit symmetry-breaking terms align the field at a
unique minimum of the vacuum manifold so that as larger scales
come into the horizon, the Higgs field is already correlated and
textures do not form.

To illustrate our point, we will first briefly review how textures work.
The simplest model assumes a global $G=SU(2)$ symmetry broken by a
complex doublet $\phi^a$ with potential $\lambda(\phi^2-v^2)^2$.  At the
earliest times the temperature $T$ is greater than $v$ and the
global symmetry is unbroken; as the Universe cools and $T$
falls below $v$ the Higgs field falls to the vacuum manifold,
but it falls to different locations on the vacuum manifold in
different causally disconnected regions of the Universe.  As the
Universe expands and these different regions come into causal
contact, the field aligns itself quite rapidly within the
horizon.  Since the homotopy group $\pi_3$ of the vacuum
manifold is nontrivial,
in becoming aligned the Higgs field may become wound around the
vacuum manifold and form a texture, an unstable topological
defect, which rapidly
collapses to a point [\kibble].  When the size of the texture has
decreased to roughly $v^{-1}$ the gradient energy pulls the Higgs field
out of the vacuum manifold, the texture unwinds and the energy
is radiated away in
Goldstone bosons which results in the astrophysically relevant
density perturbation.  As new scales come across the horizon new
knots continue to form resulting in perturbations on larger
scales.  We should remind the reader that if the symmetry $G$ is
gauged then the texture field configuration is a gauge
transformation of the vacuum having no physical consequences.

We now wish to consider the effect on the evolution of textures of
higher-dimension operators that violate the continuous global
symmetry $G$. It is widely believed that Planck-scale physics
results in the violation of all global symmetries ---  both continuous
and discrete.  Wormholes provide one specific mechanism for this violation
[\gs,\coleman,\gilbert], but for those uneasy about wormhole arguments
there is a very simple physical argument  for why one still
expects all global symmetries to be violated.  It is well known
that as a consequence of the black-hole no-hair theorems [\nohair]
the global charge of a black hole is not defined; therefore, if
in a scattering process a virtual or non-virtual black hole is
formed from an initial state of definite global charge, the
black hole will then unprejudicially decay (Hawking evaporate)
[\hawking] into final states of differing
global charge. At energies small compared to the Planck mass we can
represent these symmetry-violating effects by higher-dimension operators
in an effective theory of the light modes. On dimensional grounds,
the higher dimension operators are expected to be suppressed by the
appropriate power of the Planck mass
$$
\De L = g \Mp^{n-4} \phi^n,
\eqn\operator
$$
where we expect the coupling constant $g\sim\order(1)$.

It is easy to qualitatively see what the effect of these explicit
violations of the global symmetry on texture production is going to
be. Instead of an isopotential vacuum manifold (typically of the
form of an $S^3$ in field space), we now have a ``tipped'' manifold
with only a single true vacuum configuration. (In general a single
higher dimension operator would leave a discrete set of degenerate
vacuum states, so one might be concerned about the appearance
of domain walls when $\phi$ condenses. However, when they occur
these walls are
generally destabilized by the effects of yet higher dimension operators.
We will return to this point below.) As the temperature falls
below the $G$-symmetry-breaking value, correlation length sized
domains form, in which the order parameter is aligned.
If there were no mechanism
to rotate the order parameter in different domains
into a common direction before they come into the growing horizon
then topological defects, specifically textures, would form. However,
there {\it is} a mechanism -- the tipping of the
potential by the higher-dimension operators.

To be useful for galaxy or large-scale structure formation, the
texture scenario must have new textures still entering the horizon
at the time $\ts$ when the horizon volume contains
a galactic (or large-scale) mass. In other words, we need to
ensure that in the interval between the symmetry breaking phase
transition and $\ts$, the order parameter in the various
domains has not finished rotating to a common direction.
Since the expansion of the universe is so slow compared to particle
physics timescales, one can anticipate that this leads to a
very stringent condition on the coupling constants $g$.

As discussed above, textures arise in models where a global continuous
symmetry $G$ is broken to $H$ such that $\pi_3(G/H)$ is non-trivial.
The simplest example of this is the breaking $SU(2)\to 1$ as discussed
originally by Turok [\turok]. Rather than consider the effects of higher
dimension operators on this model, a moments thought convinces one that,
if one is only interested in order of magnitude estimates, it
is enough to consider a simpler toy model of $G=U(1)_{\rm
global}$.  Specifically, we take the Lagrangian of the model to
be
$$
L=|\p_\mu \phi|^2 - {\la\over4} (|\phi|^2-v^2)^2 + {g\over
\Mp^{2m+n-4}} |\phi|^{2m}\phi^n + h.c. - c
\eqn\lagr
$$
where $c$ is a constant chosen so that the vacuum energy is
zero.  We have taken the leading higher-dimension operator
that violates $G$ to be of dimension $2m+n$ with global charge
$n$.
If we write $\phi=\rho e^{i\th}$ then after
symmetry-breaking the equation of motion for $\th$ in an expanding
Universe is
$$
\ddot\th+3H\dot\th-\del^2\th+gn{\Mp^4 \over
v^2}\left({v\over\Mp}\right)^{2m+n} \sin n\th =0,
\eqn\eom
$$
where $H=1.66 g_\star^{1/2} T^2/\Mp=3/2t$ is the Hubble parameter in
a radiation dominated Universe and $g_\star\sim100-1000$ is the effective
number of relativistic degrees of freedom.  (In general, there
may be a term $\Gamma\dot\th$ in the equation of motion from the
coupling of $\th$ to other fields.  We will say more about this
below.)  If we define
$$
\tau={1\over g^{1/2} n}{1\over \Mp}\left({\Mp\over
v}\right)^{m+(n/2) -1},
\eqn\taueqn
$$
assume $\sin n\th\simeq n\th$, and look at the homogeneous
component, we find that the equation of motion for $\th$ is
$$
\ddot\th+{3\over2t}\dot\th+\tau^{-2}\th=0.
\eqn\nexteom
$$
At a temperature $T\sim v$ the phase transition occurs and an
initial value of $\th$ is selected.  As long as $t\lap\tau$ the
field rolls slowly toward the minimun until $t\sim\tau$ at which
point the field begins to undergo oscillations with an amplitude
which decreases as $(t/\tau)^{-3/4}$.  These coherent scalar-field
oscillations describe a condensate of zero-momentum Goldstone
bosons, and the decaying amplitude describes their dilution in
the expanding Universe.  The important point is that from
Eq.~\nexteom\ it is clear that the field becomes localized near
the minimum of the potential roughly within a few timescales
$\tau$.

For the texture model to account for the large-scale structure in the
Universe, the timescale $\tau$ for rolling must be longer than
or comparable to the time $\ts$ at which galactic (or larger) scales
enter the horizon.  Taking the smallest scale of interest
(for the most conservative limit on the coupling $g$), namely
galactic scale, that which encloses $10^{12}\,M_\odot$, we find
that $\ts\sim 4\times 10^6$ sec.  So if textures are to be
effective in forming even galactic-sized structure,
$$
7\times10^{-47}\, {1\over g^{1/2} n}(10^3)^{m+(n/2)}
\left({10^{16}\,{\rm GeV}\over v}\right)^{m+(n/2)-1}
\gap 4\times10^6.
\eqn\condition
$$
There are various ways of expressing the limits that result from
this condition.  If we demand $v\sim\order(10^{16})$ GeV, as
required for the appropriately sized density perturbations, and
suppose that the explicit breaking is due to a dimension-5
operator, then we find that the coupling is constrained to be
$$
g\lap 10^{-91}.
\eqn\gcond
$$
This is the central result of our Letter.

If instead we demand that $g\sim\order(10^{-2})$ with the same
value of $v$ then we find that first operator that explicitly
breaks the symmetry and is consistent with texture-seeded
galaxies has dimension $2m+n\simeq35$.
Finally, if we assume a
symmetry-breaking operator of dimension 5 and again take
$g\sim10^{-2}$ we find that $v\lap10^{-5}$ eV, clearly
incompatible with the texture scenario.

One might be concerned that the field $\th$ could get hung up
near $n\th=\pi$, where the potential is a maximum, in certain
regions of the Universe leading us to underestimate the rolling
timescale $\tau$.  However, horizon-sized ($\la\sim H^{-1}$) thermal
fluctuations of $\th$ at a temperature $T\sim v$ are roughly
$$
\VEV{(\delta\th)_{\la}^2}_T \sim g_\star^{1/2}
\left({v\over\Mp}\right) \sim 10^{-2},
\eqn\delth
$$
so $\delta\th\sim 10^{-1}$, and there is no problem with
the field getting hung up at its maximum.  This should be
no surprise since we know that in inflationary models the
inflaton rolls slowly only if the potential is extremely
flat.

While our analysis is quite simple and we ignore such effects as
the running of the coupling constants $g$, it is clear from
Eq.~\gcond\ that a more refined calculation would not alter
our conclusions.  We have also considered the effect of a
friction term $\Gamma\dot\th$ in the equation of motion that
could arise as result of coupling of $\th$ to other fields in
the theory.  However, since the mass of the field $\th$ is so
small, we expect $\Gamma$ to be small; furthermore,
astrophysical considerations constrain the coupling of $\th$ to
other fields in the theory to be small [\spergel].
Still, we have checked that if we take the extreme conservative
limit, that if for some unforeseen reason $\Gamma$ assumes it
maximum (in-)conceivable value $\Gamma=\Mp$, our results are
qualitatively robust (i.e. $g$ is still constrained to be
extremely small).

We should also
point out that if $\Gamma$ is very small
(as we expect), then one might worry that the energy density in
the condensate of $\th$ particles at some point might become
greater than that in radiation.  Indeed, the simple condition
that relic $\th$ particles do not ``overclose'' the Universe may
be used to constrain $g$; however, the constraint turns out to
be weaker than Eq.~\gcond\ and the assumption of a
radiation-dominated Universe in our analysis remains valid.

While discrete and continuous global symmetries are in general
explicitly violated by Planck-scale physics, this is not true of
gauge symmetries.  As discussed in Ref.~\turok, we can imagine the
necessary continuous global symmetry $G$ arising as an accidental
global symmetry of the effective Lagrangian when we impose a set
of discrete symmetries.  If these discrete symmetries are
discrete {\it gauge} symmetries [\kw,\amr] which are protected from
violation by Planck-scale physics, then not only the
renormalizable part of the effective Lagrangian, but also some
higher-dimension operators will be constrained to be invariant
under $G$ (especially if these discrete symmetries are
nonabelian [\non]).  However, it requires some stretch of the
imagination
to have a sufficiently large discrete gauge symmetry group to
forbid the existence of all higher-dimension operators up to
dimension 35, although it is a possible way of ensuring the
integrity of the texture scenario.  On the other hand, if
observers find the cosmic-microwave-background distortions predicted
by textures, we could construe this to be evidence of a very
large discrete-gauge symmetry group (as have been conjectured in
string theory [\string]).  We note that if the breaking of the
global symmetry also involves the spontaneous breaking of
discrete gauge symmetries then domain walls form that are {\it
not} destabilized by higher-dimension operators, and the
evolution of the Universe is radically different (and ruled
out).

We can also
consider the more recent models
of large-scale structure formation involving perturbations
through gradient energy without topological defects [\dst].  In
this case arguments analogous to those above show that this
scenario is also destabilized by Planck-scale effects.

In summary, we find that the couplings of higher-dimension
global-symmetry-violating terms in the effective Lagrangian must
be unnaturally small in order for textures to be responsible
for the formation of large-scale structure.  Admittedly, little
is known about Planck-scale physics; however,
the agnostics who question the existence of higher dimension
operators coming from quantum-gravity effects must nevertheless
certainly conclude that the texture model is at least unstable
to the unknown.  Our argument is
simply that unless the (approximate) vacuum manifold of $G$ is exactly flat,
the field in causally disconnected regions of the Universe will
become aligned before coming into causal contact.
In a sense the unnatural flatness of the potential necessary for
the survival of the
texture scenario echoes that required in new inflationary
models.  Namely, that the field rolls only very slowly to its
minimum.

We understand that similar conclusion have been reached by
R.~Holman, S.~Hsu, E.~Kolb, R.~Watkins, and L.~Widrow.  We
gratefully thank Jacques Distler and Sandip Trivedi
for discussions, Michael Turner and David Spergel for useful
comments, and John Preskill for encouraging us to report
our results.  MK gratefully acknowledges the hospitality of the
Institute for Advanced Study.

\par

\refout

\end